\documentstyle[12pt]{article}
\def\lsim{\mathrel{\raise.2ex\hbox{$<$}\hskip-.8em\lower.9ex\hbox{$\sim$}}}
\def\gsim{\mathrel{\raise.2ex\hbox{$>$}\hskip-.8em\lower.9ex\hbox{$\sim$}}}
\def\geq{\mathrel{\raise.2ex\hbox{$>$}\hskip-.8em\lower.9ex\hbox{$-$}}}

\begin{document}

\font\fortssbx=cmssbx10 scaled \magstep2
\hbox to \hsize{
\hskip.5in \raise.1in\hbox{\fortssbx University of Wisconsin - Madison}
\hfill$\vcenter{\hbox{\bf MAD/PH/855}
                \hbox{\bf DTP/94/110}
                \hbox{December 1994}}$ }

\vspace{.5in}

\begin{center}
{\large\bf A determination of the QED coupling at the $Z$ pole}\\[.25in]
A.D. Martin\footnote{On leave from the Department of Physics, University of
Durham, England}  and D. Zeppenfeld\\
\it Department of Physics, University of Wisconsin, Madison, WI 53706
\end{center}

\vspace{1in}

\begin{abstract}
We critically examine the evaluation of the hadronic contribution to the
running of the QED coupling, $\alpha$, from $Q^2=0$ to $Q^2=M_Z^2$. Using
data for $e^+e^-\to{}$hadrons we find that
$\alpha(M_Z^2)^{-1} = 128.99\pm 0.06$, as compared to the existing value of
$128.87\pm 0.12$.
\end{abstract}

\thispagestyle{empty}

\newpage
\setcounter{page}{1}

The improvement in the measurements of electroweak quantities allows high
precision tests of the Standard Model in which the measured $Z$ boson mass
is related to other observables (see, for example, the recent reviews in
Refs.~\cite{pdg1,rev}). Surprisingly, out of the three accurately measured
quantities ($\alpha$, $G_F$, $M_Z$) which determine the Standard Model, the
largest uncertainty comes from the running of $\alpha$ from $Q^2=0$, where it
is precisely known, up to the $Z$ pole, which is the scale relevant for the
electroweak precision tests. Indeed, other electroweak quantities are being
measured with an accuracy comparable to that associated with $\alpha(M_Z^2)$.
The source of the ambiguity in the value of $\alpha(M_Z^2)$ is the hadronic
contribution to the photon vacuum polarization $\Pi(s)$. This contribution
is determined by the dispersive sum of all possible hadronic states produced
in $e^+e^-$ annihilation into hadrons via an intermediate photon
\begin{eqnarray}
{\rm Re}\, \Pi_h(s)
&=&{\alpha s\over3\pi}P\int_{4m_\pi^2}^\infty
{R^{\gamma\gamma}(s')\over s'(s'-s)}ds'\;, \label{eq:Pi_h}
\end{eqnarray}
with $\alpha^{-1}=137.036$ and
\begin{equation}
R^{\gamma\gamma} = {\sigma(e^+e^-\to\gamma\to\rm hadrons) \over
\sigma_{\mu\mu}}\;.
\end{equation}
Here $\sigma_{\mu\mu}=4\pi\alpha(s)^2/3s$ is the lowest order point-like
$e^+e^-\to\gamma\to\mu^+\mu^-$ cross section. The $\mu\mu$ cross section
is expressed in terms of the running coupling $\alpha(s)$ in order to
eliminate any QED effects from the hadronic contribution to the
current-current two-point function $\Pi_h/\alpha$.

The currently accepted determination of the running of $\alpha$ is based on
the analysis of Burkhardt {\it et al.}~\cite{bur}. They used data for
$R^{\gamma\gamma}$, supplemented by narrow resonance contributions, to
obtain a hadronic contribution to the running of $\alpha$ of
\begin{equation}
\Delta \alpha_h(M_Z^2) = -{\rm Re}\, \Pi_h(M_Z^2) = 0.0288\pm 0.0009 \;.
\end{equation}
When the leptonic contribution,
\begin{equation}
\Delta \alpha_{\ell}(M_Z^2) = {\alpha\over 3\pi}\sum_{\ell}
\left[ {\rm ln}{M_Z^2\over m_\ell^2} - {5 \over 3} +
{\cal O}({m_\ell^2\over M_Z^2})\right] = 0.03142 \;,
\end{equation}
is added, their result gives
\begin{equation}
-{\rm Re}\, \Pi(M_Z^2) \equiv \Delta \alpha
= \Delta\alpha_h(M_Z^2) + \Delta\alpha_\ell(M_Z^2) = 0.0602\pm 0.0009 \;,
\label{eq:alpha}
\end{equation}
which  translates to
\begin{equation}
\alpha(M_Z^2)^{-1}= (1-\Delta \alpha)\alpha^{-1}
 = 128.78\pm 0.12 \;.
\end{equation}
The analysis\cite{bur} was subsequently updated by Jegerlehner~\cite{jer} to
give
\begin{eqnarray}
\Delta\alpha_h(M_Z^2) & = & 0.0282 \pm 0.0009\;, \\
\alpha(M_Z^2)^{-1} & = & 128.87 \pm 0.12\;.
\end{eqnarray}
Here we are using the effective QED coupling, which is denoted by $\bar\alpha$
in the {\it Review of Particle Properties}~\cite{pdg1,pdg}.

For electroweak precision tests it is important to see if the determination
of ${\rm Re}\, \Pi_h(M_Z^2)$ (and hence of $\alpha(M_Z^2)$) can be improved
and, in particular, the error reduced. In the following we compare our
analysis with the original work of Ref.~\cite{bur} since it lists the
various contributions to $\Delta\alpha_h$ in detail. Indeed the first
column of Table~\ref{table:contrib}, which is taken directly from
Ref.~\cite{bur}, shows the contributions to ${\rm  Re}\, \Pi_h$ from
different $W\equiv\sqrt {s'}$ regions of the dispersion integral, together
with the associated errors. We note that the largest error arises from
the region $2.3<W<9$~GeV, which contains the $c\bar c$ resonance region
together with the `continuum' regions below both the $c\bar c$ and the
$b\bar b$ threshold. The evaluation of the contributions from this
region, used in Ref.~\cite{bur}
\footnote{The updated analysis~\cite{jer} uses data from the Crystal Ball
collaboration~\cite{CB}.},
relied on the original MARK\,I~\cite{mark} data for $R(W)$.
In fact the major uncertainty in the determination of Re\,$\Pi_h$ is
associated with the normalization errors of the measurements of
$\sigma(e^+e^-\to{}$hadrons). Thus the 10\% error associated with
the $2.3<W<9$~GeV contribution of Ref.~\cite{bur} reflects the 10\%
normalization uncertainty of the MARK\,I data.

In the continuum regions well above the $q\bar q$ thresholds we are now
in a position to use perturbative QCD to predict $R^{\gamma\gamma}$
extremely accurately. First our knowledge of the QCD coupling has
considerably improved since the previous calculation~\cite{bur} of
Re\,$\Pi_h$. Even if we take a conservative view of the average of all
of the measurements of $\alpha_s$~\cite{pdg}, we can conclude
$\alpha_s(M_Z^2)=0.118\pm0.007$. Moreover $R^{\gamma\gamma}$ is known
up to, and including, the ${\cal O}(\alpha_s^3)$ contributions, and
the running of $\alpha_s$ is known to 3 loops. Finally we now have
experimental evidence of the value of the top quark mass $m_t$. Thus,
for example, if we evaluate $R^{\gamma\gamma}$ at values of $W=3,\ 9$
and 150~GeV just below the $c\bar c, b\bar b$ and $t\bar t$ thresholds
respectively, we find
\newpage
\begin{eqnarray}
R^{\gamma\gamma} &=& 2.17\pm 0.03 \qquad \mbox{at $W=3$ GeV}\;,
\label{eq:W=3}\\
                 &=& 3.58\pm 0.04 \qquad \mbox{at $W=9$ GeV}\;,\\
                 &=& 3.80\pm 0.01 \qquad \mbox{at $W=150$ GeV}\;.
\label{eq:W=150}
\end{eqnarray}
We allow for the change in the number of flavors at each $q\bar q$
threshold both in $\alpha_s(s)$ and in $R^{\gamma\gamma}(s)$. The errors
include the $\pm 0.007$ uncertainty in the input value of $\alpha_s(M_Z^2)$,
an uncertainty from yet unknown ${\cal O}(\alpha_s^4)$ terms (taken to be
equal in size to the $\alpha_s^3$ contribution to $R^{\gamma\gamma}$), and
uncertainties from threshold effects. The main threshold uncertainties
arise in the $b\bar b$ channel because we combine data on the resonance
contributions with the QCD formula. We estimate these effects by comparing
a naive $\beta(3-\beta^2)/2$ threshold behaviour with the full
${\cal O}(\alpha_s)$ QCD formula~\cite{schwinger} and find that
$b\bar b$ threshold uncertainties contribute very little to the error
on $\alpha(M_Z^2)$.

Given that $R^{\gamma\gamma}$ is known so precisely in the continuum
regions, we may use it to improve the normalization of the experimental
measurements of $\sigma(e^+e^-\to{}$hadrons). Such a program was
carried out for 21 experiments by Marshall~\cite{marsh} in a detailed
study performed in 1988. He noted that many experiments which partially
overlap the MARK\,I region have smaller systematic errors than the
MARK\,I data. As a result of a global QCD fit to the world data for
$R$ he concluded that the experimental normalization of $1.00\pm0.10$
of the MARK\,I data should be adjusted to $0.850\pm0.019$ to bring
them into line with the world data (which was by far the biggest adjustment
of data that he obtained). In their paper~\cite{mark} MARK\,I quote a
normalization error of $\pm20\%$ at $W=2.6$~GeV decreasing smoothly to
$\pm10\%$ for $W>6$~GeV. Marshall's adjustment brought the MARK\,I data
into excellent agreement with QCD expectations in the continuum regions
below the $b\bar b$ and $c\bar c$ thresholds. Clearly such an adjustment
will have a dramatic effect on the value, and the accuracy, of
$\alpha(M_Z^2)$. Incidentally at the time of Marshall's analysis
the coefficient of the $(\alpha_s/\pi)^3$ term in $R^{\gamma\gamma}$
was erroneously large and of the wrong sign (+65 instead of $-$12.8 for
five flavors). As a consequence the renormalization factors found by Marshall
should have been even smaller.

To evaluate (\ref{eq:Pi_h}) we assume that $R^{\gamma\gamma}$ is given by
perturbative QCD in the continuum regions $3<W<3.9$~GeV and $6.5<W<\infty$,
apart from the $\psi$ and $\Upsilon$ resonance contributions. Typical errors
on $R^{\gamma\gamma}$ in these regions are shown in
Eqs.~(\ref{eq:W=3})--(\ref{eq:W=150}). Following Marshall's procedure, we
also use the continuum values of $R^{\gamma\gamma}$ to normalize the various
data sets. For the MARK I data~\cite{mark} we find that an overall
renormalization of $0.83 \pm 0.02$ is required. In fact fitting to the
MARK~I data in the $W<3.85$~GeV continuum region gives  essentially the
same renormalization factor as fitting to their  $W>6.5$~GeV
continuum data, see Fig.~\ref{fig:R^gg}(a). One option would be to
use the renormalized
MARK I data to evaluate the contribution to (\ref{eq:Pi_h}) from the
interval $3.9<W<6.5$~GeV, between the two continuum regions. To be
precise we could use the curve of Fig.~\ref{fig:R^gg}(a) in which the portion
between $4.5<W<6.5$~GeV is shown dotted. To check this `MARK I' curve,
we compare
with the more recent and precise Crystal Ball measurements~\cite{CB,CB1}.
First we renormalize the Crystal Ball ('90) measurements~\cite{CB}
using their
$W>6.5$~GeV data. A factor $1.06\pm0.02$ is found. Indeed with this
renormalization all the Crystal Ball ('90) data are well described by
$R$(QCD), see Figs.~\ref{fig:R^gg}(a,b). From Fig.~\ref{fig:R^gg}(a)
we see that the more precise
Crystal Ball ('90) data lie significantly above the renormalized MARK I data
for $W\approx 5$~GeV. Evidence in favor of the higher Crystal Ball values
and, in particular, for taking the continuous (rather than the dotted) curve
for $R$, comes from two first-generation experiments, DASP~\cite{dasp} and
PLUTO~\cite{pluto}, see Figs.~\ref{fig:R^gg}(c,d) respectively. The data
are shown after they have been renormalized
\footnote{We obtain similar normalization factors to Marshall~\cite{marsh},
after allowing for the change in the ${\cal O}(\alpha_s^3)$ contribution.}
to $R$(QCD) at $W\approx 3.6$~GeV.  On the other hand for $W<4.5$~GeV the
data of Figs.~\ref{fig:R^gg}(b,c,d) show that the line drawn
through the MARK I
data is a reasonable representation of $R$ in the $c\bar c$ resonance region.
We note that both the Crystal Ball ('86)~\cite{CB1} and the DASP~\cite{dasp}
data are able to resolve the $\psi (4.04)$ and $\psi(4.16)$ states and,
especially, that our curve is a good average of $R$ for this resonance
region. The contributions to Re\,$\Pi_h$, obtained from integrating
$R$ over the continuous curve in Fig.~\ref{fig:R^gg}, are shown in
Tables~\ref{table:contrib} and \ref{table:summary}, together with the
contributions of the  families of $\psi$ and $\Upsilon$ resonances. The
resonance contributions are determined from~\cite{bur}
\begin{equation}
\Pi_{\rm res} = -{3\Gamma_{ee}\over M} \,
{\alpha\over\alpha\left(M^2\right)^2} \;, \label{Pi_res}
\end{equation}
where $M$ and $\Gamma_{ee}$ are the mass and leptonic width of the
resonance, respectively, and where we account for the running of the
effective QED coupling at the resonance scale. Equation~(\ref{Pi_res})
follows from integration over a narrow Breit-Wigner resonance form.

The errors associated with the continuum contributions are estimated as for
Eqs.~(\ref{eq:W=3})--(\ref{eq:W=150}). For the intervening interval,
$3.9<W<6.5$~GeV, we estimate the error by repeating the calculation using
the dotted line in Fig.~\ref{fig:R^gg}(a). The contribution reduces from
$2.90\times  10^{-3}$ to $2.71\times 10^{-3}$, and we regard this change
as representative of the uncertainty of this interval.

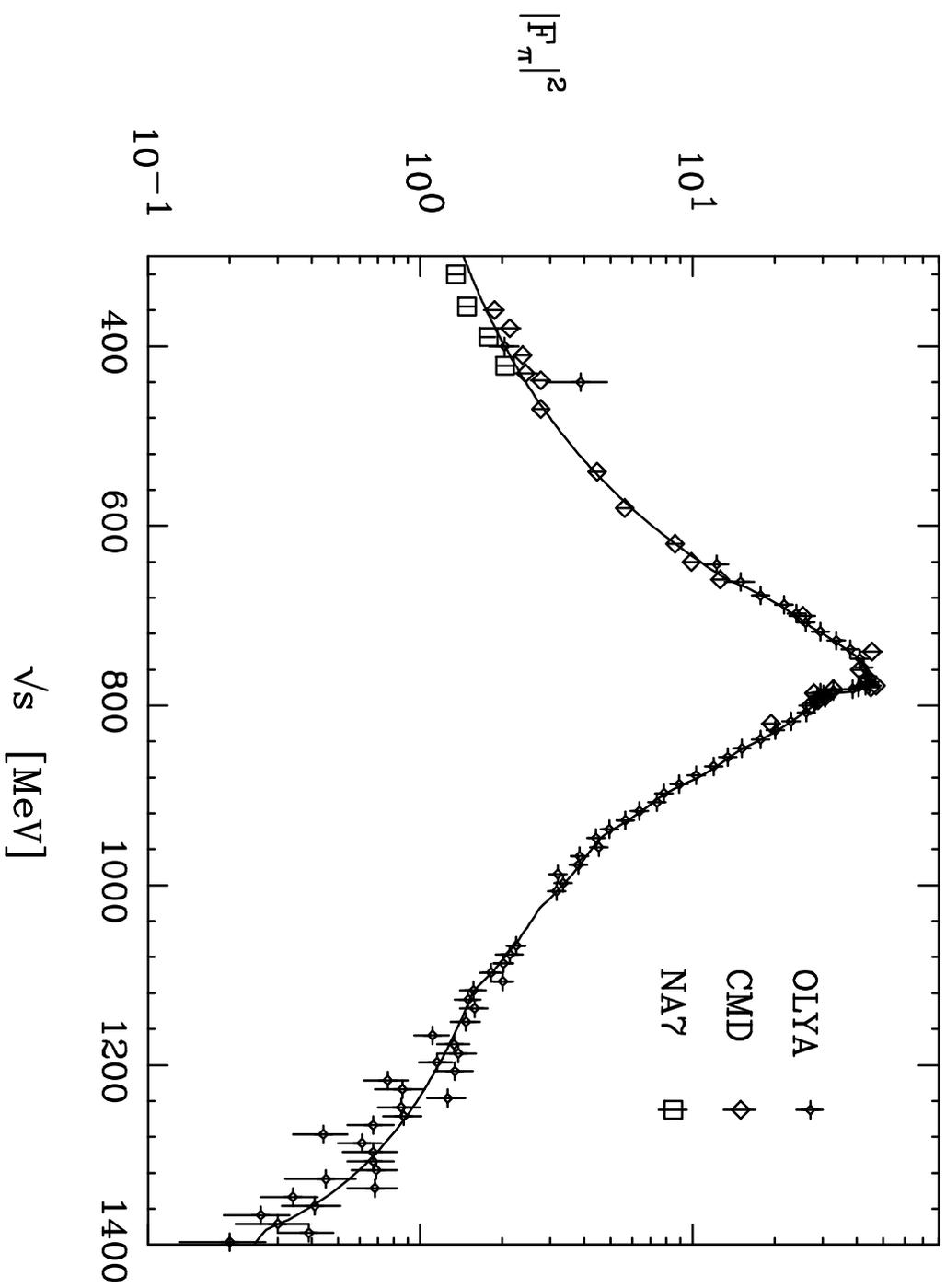
\begin{figure}[h]

\caption{\label{fig:rho}
Data on the pion form factor~\protect\cite{pi2}. The curve is the
representation of the data which is used to evaluate the $\pi\pi$
contribution.
}
\end{figure}

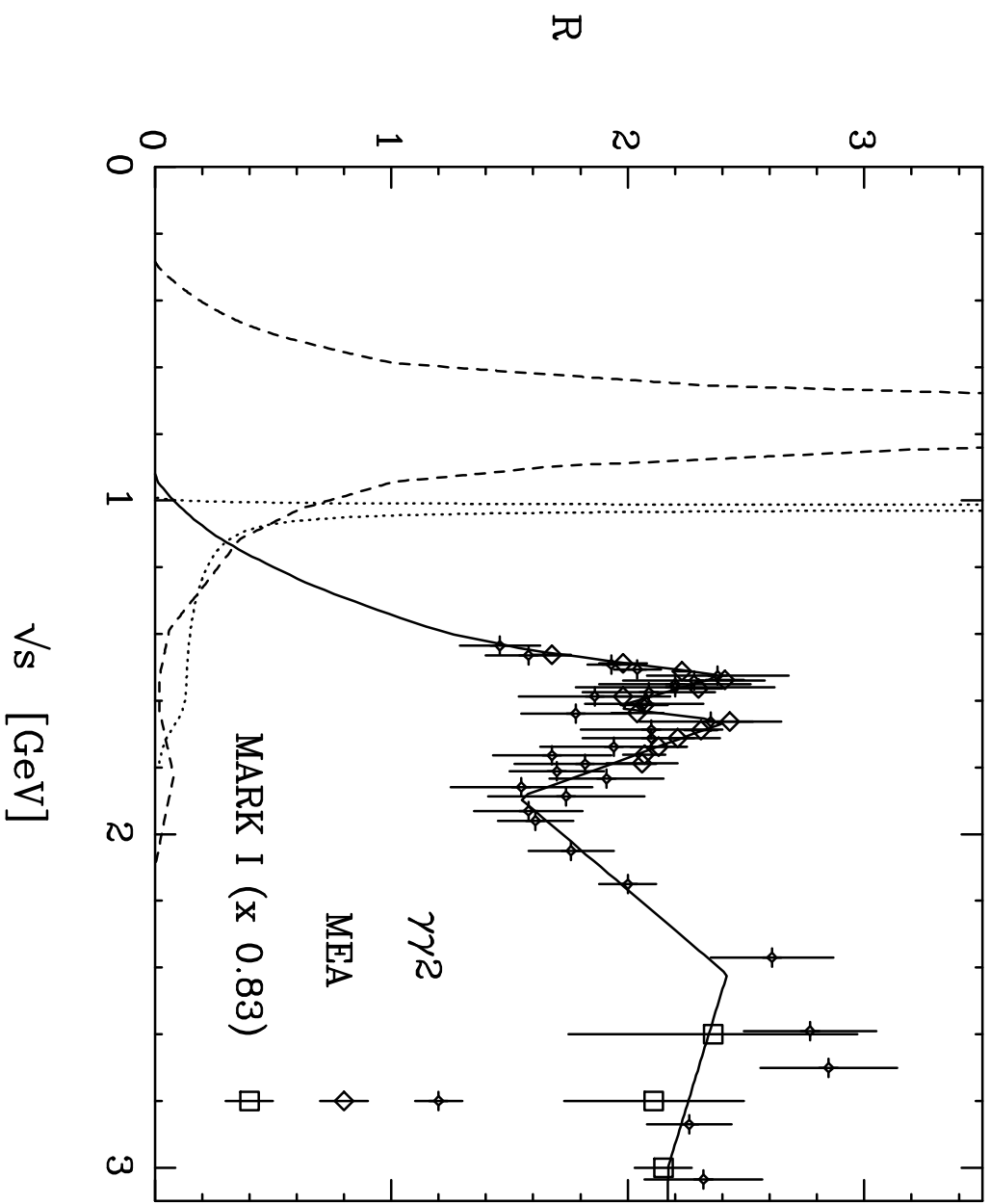
\begin{figure}[h]

\caption{\label{fig:R}
The dashed and dotted curves are a representation of the data for the
contribution to $R$ from $e^+e^-\to \pi^+\pi^-$ and $K\bar K$ respectively.
The continuous curve represents the data for multi ($\geq 3$) pion production.
For $W<1.45$~GeV it represents the sum of the multipion exclusive channels.
Above $W=1.45$~GeV the multipion data are taken from
Refs.~\protect\cite{mark,pi3}. Also shown are three MARK\,I
measurements~\protect\cite{mark}, renormalized by a factor of 0.83.
We numerically integrate over the $\rho$ and $\phi$ resonance forms,
but include the $\omega(782)$ contribution via (\protect\ref{Pi_res}).}
\end{figure}

We now turn to the region below $W=3$~GeV. Here there are many experiments
measuring $e^+e^-$ annihilation to specific hadronic channels. We evaluate
the contribution to Re\,$\Pi_h$ from this region in four separate parts;
see Table~\ref{table:summary}. First we calculate the contribution
from $e^+e^-\to\pi^+\pi^-$ by integrating over $R$ obtained from detailed
measurements~\cite{pi2} of the pion form factor $F_\pi(s)$ via
\begin{equation}
R(s) = {1\over 4}\left(1-{4m_\pi^2\over s}\right)^{3/2}
\left| F_\pi(s) \right|^2 \;.
\end{equation}
To be specific we integrate over the curve shown in Fig.~\ref{fig:rho}
and assign to this contribution a $\pm4\%$ uncertainty arising
primarily from the experimental normalization of
$|F_\pi (s)|^2$. This region is dominated by the $\rho$ resonance (with the
resonance shape mutilated by $\rho$--$\omega$ interference). For example
the intervals $1<W<1.4$ and $1.4<W<2$~GeV only give contributions to
$-{\rm Re}\, \Pi_h$ of $0.16\times 10^{-3}$ and $0.02\times 10^{-3}$
respectively. Secondly, we include the contribution
due to the $\omega(782)$ resonance using (\ref{Pi_res}), where the
error reflects the uncertainty in the observed leptonic width~\cite{pdg}.
Thirdly, we calculate the $e^+e^-\to K\bar K$ contribution to Re\,$\Pi_h$
using the parametric form for the kaon form factor determined by Bisello
{\it et al.}~\cite{bis} from a fit to $e^+e^-\to K^+K^-$ data~\cite{bis,kk}.
By far the dominant contribution here comes from the $\phi$ resonance,
though there are small contributions from the $\rho,\omega\to K\bar K$
resonance tails and an even smaller contribution from the 1.5--1.6~GeV
resonance region; see the dotted curve in Fig.~\ref{fig:R}. The error
reflects the uncertainty in the $\phi$ leptonic width~\cite{pdg}. To a
good approximation the total contribution is twice the $K^+K^-$
contribution. On the $\phi$ resonance this allows for $K^0\bar K^0$
and $3\pi$ contributions, while above the resonance it represents the
total $K\bar K$ contribution since, away from threshold, the effect of
the  $K^+K^- - K^0\bar K^0$ mass difference is suppressed. Finally we
have the contributions to Re\,$\Pi_h$ from the region above $W=0.9$~GeV
due to multi ($\geq 3$) pion
production. For the region $0.9<W<1.45$~GeV we use the sum of the data
for the exclusive channels~\cite{pi}. The dominant contribution comes from
$\pi^+\pi^-\pi^0\pi^0$ and $\pi^+\pi^-\pi^+\pi^-$ production.
Above $W=1.45$~GeV we use a line through the $R$ data of
Refs.~\cite{pi3}, joining on to the QCD value at $W=3$~GeV, as shown
in Fig.~\ref{fig:R}.

It is instructive to compare our results with those of Ref.~\cite{bur}.
{}From Table~\ref{table:contrib} we see that the main improvement in
accuracy comes from the $2.3<W<9$~GeV and $12<W<\infty$ regions, due
mainly to our use of $R^{\gamma\gamma}$(QCD). The difference in size of
the $2.3<W<9$~GeV contribution can be attributed to our use of renormalized
MARK~I and Crystal Ball '90 data. To make an
exact comparison with Ref.~\cite{bur} we should let $m_t\to\infty$, rather
than taking the value $m_t=174$~GeV that we have used to include the
$e^+e^-\to t\bar t$ contribution. The resulting effect is that
$-{\rm Re}\, \Pi_h(M_Z^2)$ would increase very slightly to
$27.39\times10^{-3}$.

Although we have numerically integrated over $\rho$ and $\phi$ resonant
shapes, and carefully considered individual contributions to $e^+e^-$
production processes, we find that there is a comparatively modest
reduction in the error of the contribution to Re\,$\Pi_h$ from the
low energy, $W<3$~GeV, region. From the first five rows of
Table~\ref{table:contrib} we find that the contribution for
$W<2.3$~GeV is $-(6.06\pm0.25)\times10^{-3}$ as compared to
$-(6.34\pm0.43)\times10^{-3}$ of the previous calculation~\cite{bur}.
In fact we see from Table~\ref{table:summary} that the error
$\pm0.33\times10^{-3}$ on the $W<3$~GeV contribution limits the
present accuracy of $\alpha(M_Z^2)$.

In conclusion we find that
\begin{equation}
\alpha(M_Z^2)^{-1} = 128.99\pm0.06 \;.
\end{equation}
The large reduction in the error, in comparison to that found in
Refs.~\cite{bur,jer}, arises because we use the precise perturbative QCD
prediction for $R^{\gamma\gamma}$ in the regions well above $q\bar q$
thresholds. As a consequence we find (i)~that the contribution from the
high $W\equiv\sqrt s$ region is well determined, and (ii)~that we are
able to reliably normalize the data in the $c\bar c$ resonance region
which significantly reduces the value {\em and the uncertainty} of
$\alpha(M_Z^2)$. Although perturbative QCD stabilizes the contribution
from the $W\gsim3$~GeV region, the contribution from lower values of
$W$ relies directly on the available data for $e^+e^-$ annihilation
into hadrons. In fact the dominant uncertainty arises from the $1<W<3$~GeV
region and, in view of the importance of an accurate value of $\alpha(M_Z^2)$
for precision tests of the Standard Model, it is crucial to improve the
accuracy of the low energy measurements of $\sigma(e^+e^-\to{}$hadrons).

\bigskip
\leftline{\bf Acknowledgments}

We thank K.~Hagiwara for valuable discussions, and M.G.~Olsson and
M.R.~Whalley for advice concerning the data. The data used in this analysis
can be found in the Durham-RAL High Energy Physics database.
This research was supported in
part by the University of Wisconsin Research Committee with funds granted
by the Wisconsin Alumni Research Foundation, by the U.~S.~Department
of Energy under contract No.~DE-AC02-76ER00881, and in part
by the UK Particle Physics and Astronomy Research Council.

\bigskip
\noindent
{\bf Note added:} After this work was completed we received a paper
entitled ``Re-evaluation of the Hadronic Contribution to
$\alpha(M_Z^2)$" by M.~L.~Swartz, SLAC-PUB-6710, November~1994,
in which he obtained $\Delta\alpha_h(M_Z^2)=(26.66 \pm 0.75)\times10^{-3}$,
and hence $\alpha(M_Z^2)^{-1}=129.08\pm0.10$. We note that the
error estimates would be expected to differ, since the two
(completely independent) calculations differ in their reliance on $R$(QCD).

\newpage

\begin{table}
\caption{Contributions to $-1000\,{\rm Re}\,\Pi_h(M_Z^2)$.
\label{table:contrib}}
\medskip
\centering
\begin{tabular}{|c|rcl|rcl|}
\hline
$W$ range (GeV)& \multicolumn{3}{c|}{Burkhardt et al.\ \cite{bur}}&
\multicolumn{3}{c|}{This work}\\
\hline
$2m_\pi$--1.0& \multicolumn{3}{c|}{---}& 3.21& $\pm$& 0.14\\
$\rho$& 3.484& $\pm$& 0.171& \multicolumn{3}{c|}{---}\\
$\omega$& 0.347& $\pm$& 0.021& 0.31& $\pm$& 0.01\\
1.0--2.3& 1.981& $\pm$& 0.391& 1.95& $\pm$& 0.21$^\ddagger$ \\
$\phi$& 0.528& $\pm$& 0.024& 0.59& $\pm$& 0.02$^\dagger$\\
2.3--9.0& 7.208& $\pm$& 0.721& 6.56& $\pm$& 0.22$^\ddagger$ \\
$\psi$'s& 1.084& $\pm$& 0.057& 0.92& $\pm$& 0.05\\
9.0--12.0& 1.686& $\pm$& 0.169& 1.67& $\pm$& 0.06$^*$ \\
$\Upsilon$'s& 0.118& $\pm$& 0.005& 0.12& $\pm$& 0.01\\
12.0--$\infty$& 12.368& $\pm$& 0.371& 11.99& $\pm$& 0.09$^*$ \\
\hline
Total& 28.8\phantom{00}& $\pm$& 0.9$^{\dagger\dagger}$& 27.32& $\pm$& 0.42 \\
\hline
$\alpha(M_Z^2)^{-1}$& 128.78& $\pm$& 0.12$^{\dagger\dagger}$&
128.99& $\pm$& 0.06\\
\hline
\end{tabular}
\end{table}

$^\dagger$Includes the full $e^+e^-\to K\bar K$ contribution.

$^*$These errors have to be added linearly.

$^\ddagger$Part of these errors have to added linearly and a part in
quadrature.

$^{\dagger\dagger}$An updated analysis~\cite{jer} gives
$-1000{\rm Re}\, \Pi_h(M_Z^2)=28.2\pm 0.9$ and \newline
\indent $\alpha(M_Z^2)^{-1}=128.87\pm 0.12$.

\newpage

\begin{table}
\caption{Summary of the evaluation of Re\,$\Pi_h(M_Z^2)$
\label{table:summary}}
\medskip
\centering
\begin{tabular}{|c|l|rcl|}
\hline
$W$ range (GeV)& \multicolumn{1}{c|}{Information used}&
\multicolumn{3}{c|}{$-1000\,{\rm Re}\,\Pi_h(M_Z^2)$}\\
\hline
$2m_\pi$--3.0& & & & \\
\multicolumn{1}{|l|}{(i) $\pi^+\pi^-$ ($W<2$)}&
Data for $|F_\pi|^2$ & 3.39& $\pm$& 0.14\\
\multicolumn{1}{|l|}{(ii) $\omega\to3\pi$}& Narrow resonance formula&
0.31& $\pm$& 0.01\\
\multicolumn{1}{|l|}{(iii) $K\bar K$ inc.\ $\phi$}&
Parametrization of $e^+e^-\to K\bar K$ data& 0.59& $\pm$& 0.02$^\dagger$ \\
\multicolumn{1}{|l|}{(iv) \vtop{\hbox{$e^+e^-\to{}$hadrons\vphantom y}
\hbox{excluding $\pi\pi$, $K\bar K$}}}& $R$(data), $e^+e^-\to 4\pi$ etc.
& 2.72& $\pm$& 0.30 \\
 & & & & \\
3.0--3.9& $R({\rm QCD})$& 0.88& $\pm$& 0.01 \\
        & $+J/\psi,\ \psi',\ \psi(3.77)$& 0.92& $\pm$& 0.05\\
3.9--6.5& Renormalized $R$(data)& 2.90& $\pm$& 0.19 \\
6.5--$\infty$& $R(\rm QCD)$& 15.49& $\pm$& 0.15 \\
        & $+\Upsilon(nS),\ n=1,\dots,6$& 0.12& $\pm$& 0.01\\
\hline
Total& & 27.32& $\pm$& 0.42 \\
\hline
\end{tabular}
\end{table}
$^\dagger$Includes the $\phi\to3\pi$ contribution.

The errors are added in quadrature, except those for $R$(QCD).

\newpage

\leftline{\bf Figure Captions}

\begin{figure}[h]

\caption{\label{fig:R^gg}
The (a) MARK I~\protect\cite{mark}, (a,b) Crystal Ball '90~\protect\cite{CB},
(c) DASP~\protect\cite{dasp}, and (d) PLUTO~\protect\cite{pluto} data for $R$,
respectively renormalized by 0.83, 1.06, 0.94, and 0.96, -- factors which
are found by fitting to $R^{\gamma\gamma}$(QCD) in the continuum regions
$W>6.5$~GeV and $W<3.9$~GeV. The Crystal Ball '86 data~\protect\cite{CB1}
in (b) are not renormalized. The continuous curve is the same in each plot
and is the representation of the data used to calculate $\alpha(M_Z^2)$.
The dotted curve is a representation of the MARK I data for $4.5<W<6.5$~GeV
which is used to estimate the uncertainty. The $J/\psi$, $\psi'$ and
$\psi(3.77)$ contributions are included using (\protect\ref{Pi_res}).}
\end{figure}

\end{document}